\documentclass[hyper,a4paper,11pt,notoc]{JHEP3}

\usepackage{epsfig,bbm,amssymb}

\newcommand{\be}{\begin{equation}}
\newcommand{\ee}{\end{equation}}
\newcommand{\ba}{\begin{eqnarray}}
\newcommand{\ea}{\end{eqnarray}}
\newcommand{\dslash}{\partial\hspace{-0.2truecm}/}
\newcommand{\pbp}{\bar{\psi}\psi}

\newcommand{\psic}{\psi^c}
\newcommand{\psicl}{\psi^c_L}
\newcommand{\psicr}{\psi^c_R}
\newcommand{\psicbar}{\bar{\psi^c}}
\newcommand{\psiclbar}{\bar{\psi^c_L}}
\newcommand{\psicrbar}{\bar{\psi^c_R}}

\title{Majorana neutrinos from warped extra dimensions}

\author{{Klaas Brakke and Elisabetta Pallante}
\\Centre for Theoretical Physics, University of Groningen, 
Nijenborgh 4, 9747AG Groningen, The Netherlands\\ E-mail: 
\email{klaas.brakke@student.rug.nl}\, \email{e.pallante@rug.nl} }

\abstract{Field theory in the presence of extra dimensions has offered interesting 
solutions to the gauge hierarchy problem, and inspired many attempts to understand the 
fermion mass generation mechanism. We study the behaviour of a bulk fermion field 
subject to non-standard boundary conditions under $Z_2$ symmetry in a slice of 
$\mbox{AdS}_5$. These {\em pseudo-Majorana} boundary conditions relate the 5d spinor 
fields to their 5d charge conjugate and generate Majorana spinors in four-dimensions. 
There is a zero mode localized on the visible brane and the masses of Kaluza-Klein 
excitations are naturally of order the weak-scale. We make connection to neutrino 
phenomenology. We also demonstrate that, contrary to previous claims, a bulk singlet 
scalar field cannot provide a successful mechanism to generate a suppressed Majorana 
mass for Standard Model neutrinos localized on the visible brane. }

\keywords{Physics beyond the Standard Model, Neutrino physics, Extra dimensions, 
Majorana spinors}

\begin{document} 
\section{Introduction}

\noindent 
Theories with extra spatial dimensions have received great attention during the past decade, 
offering plausible solutions to the gauge hierarchy problem \cite{RS}, and inspiring
 many attempts to understand the fermion mass generation mechanism.

\noindent
Large flat extra dimensions \cite{ADD} can provide a suppression mechanism of fermion 
masses via a reduction of Yukawa couplings by a volume factor $1/\sqrt{V}$, where $V$ is 
the volume of the compact extra dimensional space \cite{ADD}. Many attempts have been 
considered in this context to generate small masses for the Standard Model neutrinos, 
see e.g. \cite{ADDR, DS, MP}. 

\noindent
 Interesting alternatives arise when a non-factorizable geometry is considered 
\cite{RS, Gob, RS2} .In this case the background metric entangles the four dimensional 
space to
 the extra dimensions via ``warp'' factors $\sim e^{-k|y|}$ induced by the (constant) 
curvature $k$ of the extra spatial dimensions $y$.
The role of the volume of large flat extra dimensions is now played by the warp factor,
 without implying stringent constraints on the size of the extra dimension.
A particularly successful construction has been proposed in \cite{RS} where an 
$\mbox{AdS}_5$ space has been considered with the fifth 
dimension $y$ circularly compactified on a  radius $R$ and orbifolded to 
$S^1/{Z}_2$. The orbifold fixed points $y=0$ and $y=\pi R$ can be identified with 3-branes 
with positive and negative tension: the Planck (or invisible) brane at $y=0$ and the TeV
 (or visible) brane at $y=\pi R$. Any fundamental mass parameter $m_0$ in the original 
theory will be
 converted to an effective mass parameter $m \sim m_0 e^{-k\pi R}$ on the visible brane, 
via the warping action of the slice of $\mbox{AdS}_5$ between the two branes. 
For $kR\sim 11$ all masses of particles confined to the visible brane are naturally of 
order the weak scale.  

\noindent
One of the long standing questions in particle physics remains the one of finding a
 mechanism that naturally explains the apparent hierarchy of fermion masses in the Standard
 Model and why neutrino masses are much smaller than all charged lepton masses. 
The addition of extra dimensions offers many new possibilities, along with new challenges: 
the suppression of additional operators might become a far more complicated task in an extra
 dimensional framework. Any plausible mechanism of generation of observed phenomena, must 
also guarantee a suppression of processes like proton decay, flavour changing neutral
 currents and lepton flavour violation.

\noindent
Many interesting attempts have been made along the years in order to explain the 
smallness of neutrino masses with mechanisms induced by the presence of extra dimensions.
 This has been
 done by adding fermion fields to the bulk and maintaining their Dirac nature \cite{GN}, or 
by adding a Majorana mass term \cite{HuberShafi} or considering higher dimensional Majorana 
couplings \cite{HuberShafi_2}, or adding extra scalar fields to induce a small Majorana mass
 on the visible brane \cite{Chen}. The dynamics of bulk fermion, scalar and vector fields 
has been 
extensively considered in a supersymmetric context in e.g. \cite{GP,GP_2}. 

\noindent
In this paper we consider a particular mechanism to generate a Majorana spinor on the
 visible brane by imposing a $Z_2$ transformation property of the bulk fermion fields 
which mixes the 
fields and their charge conjugates, analogously to the Majorana condition in four 
dimensions. This boundary condition is tailored for neutrinos, and it is the generalization
 to a warped geometry of the boundary conditions first proposed in \cite{flat} in the case
 of flat extra dimensions. We show how these 
{\em pseudo-Majorana} boundary conditions far more constrain the Kaluza-Klein spectrum 
with respect to the ordinary $Z_2$ boundary conditions when reducing from five to four
 dimensions. 

\noindent
The paper is organized as follows. In Section 2 we describe the framework,  impose the
 {\em pseudo-Majorana} boundary conditions and derive the Kaluza-Klein spectrum of the
 theory. Here
 we build the connection between the 5D and the 4D charge conjugation operators and we show
 that 
the boundary conditions generate one Majorana spinor with peculiar properties induced by 
the warped geometry. In Section 3 we attempt some more phenomenological considerations,
 identifying the bulk Majorana spinor with a sterile neutrino or with a Majorana Standard 
Model neutrino.
In Section 4 we summarize alternative attempts to generate small Majorana masses and we 
reconsider the derivation of ref.~\cite{Chen}. Our conclusions are in Section 5.

\section{Bulk fermions with pseudo-Majorana boundary conditions}

\noindent
We consider a scenario with one extra dimension as first introduced in ref.~\cite{RS}. 
The fifth dimension $y$ is circularly compactified on a radius $R$ and orbifolded to 
$S^1/{Z}_2$ on the 
interval $-\pi R \le y \le \pi R$. The background metric is non-factorizable and describing 
a slice of $\mbox{AdS}_5$. The orbifold fixed points are $y=0$ and $\pi R$ which can 
be naturally 
intepreted as the location of two 3-branes, where the stress-energy tensor solution of 
the Einstein's equation is non zero. 
The Planck scale brane is located at $y=0$ and the Tev or visible brane is located 
at $y=\pi R$. 
This thin-brane case can also be understood as the limiting behaviour of a thick 
domain-wall solution.

\noindent
The action for a spin-1/2 Dirac fermion $\psi$ living in the bulk is the Dirac action 
generalized to curved five-dimensional space with constant curvature and can be written as 
follows
\be
\label{eq:action_5d}
S= \int\, d^4x \int\, dy\, \sqrt{-g}\, \left ( i\bar{\psi} g^{MN}\gamma_M\,D_M \psi - 
m \bar{\psi}\psi \right )\, ,   
\ee   
where $g=det(g_{MN})= -e^{-8\sigma}$ is the determinant of the metric tensor 
with $\sigma = k|y|$ and $k$ the constant curvature along the fifth dimension. 
We use capital indices $M,N\ldots$ for objects defined in curved space and lower-case greek 
indices $\mu ,\nu,\ldots \alpha ,\beta$ for objects defined in the 
tangent frame: in our notation $\mu = 0,1,2,3$ and $M = (\mu , 5)$.

\noindent
A line-element solution of the Einstein's equations in five-dimensions, which respects 
four-dimensional Poincare invariance, is given by \cite{RS} 
\be
\label{eq:ds2}
ds^2 = e^{-2\sigma}\eta_{\mu\nu}dx^\mu dx^\nu +dy^2\, ,
\ee
 with flat metric tensor $\eta_{\mu\nu} = (-1, 1, 1 , 1)$ and $\mu ,\nu = 0,1,2,3$. 
The warp factor $e^{-2\sigma}$ 
entangles the four dimensional space-time to the fifth dimension $y$ 
rendering the geometry non-factorizable. This is a special case of an $\mbox{AdS}_5$
 geometry.
This solution can only be trusted if $k<M$, i.e. the bulk curvature is small compared with 
the fundamental Planck scale.
The gamma matrices $\gamma^M$ are the curved space Dirac matrices, $\gamma^M = 
(\gamma^\mu , \gamma_5)$\footnote{We use the Dirac representation of the gamma 
matrices unless differently stated}. They are related  to 
the flat space ones through the {\emph {vielbein}} $e^M_\alpha$ , 
$\gamma^M = e^M_\alpha\gamma^\alpha$ with $e^M_\alpha = (e^\sigma ,  e^\sigma ,
 e^\sigma , e^\sigma , 1)$.

\noindent
We have introduced the covariant derivative acting on the fermion field in curved space
 $D_M = \partial_M + \Gamma_M$, where $\Gamma_M$ is the spin connection\footnote{This is 
in agreement with ref.~\cite{GP} and one can easily check this to be equivalent to the 
prescription derived in ref.~\cite{GN}.}  
\be
\Gamma_\mu = \frac{1}{2} \gamma_5\gamma_\mu {d\sigma\over dy}\qquad\Gamma_5 =0\, . 
\ee
For the sake of simplicity, we do not consider a bulk gauge field coupled to the bulk 
fermion in this paper. An investigation of this case for flat extra dimensions can be
 found in ref.~\cite{flat}. We have introduced a
bulk mass term with mass $m$ for the fermion field. 
The mass $m$ can be typically generated by scalar fields acquiring a vacuum expectation 
value in the higher dimensional theory. 
It is in general complex valued $m= |m|e^{i\alpha}$, however its phase is irrelevant at 
this level, since we can always rotate the Wyel spinors by a phase which
 cancels the phase in the mass term and leaves the action invariant. Hence, we can set 
$m=|m|$ without loss of generality.  

\noindent
The invariance of the bulk mass term under the ${Z}_2$ parity transformation 
$y \to -y$ can be 
guaranteed in two ways: the mass $m$ and the scalar fermion bilinear $\pbp$ 
are both odd or both even under a ${Z}_2$-parity transformation. A $Z_2$ odd mass parameter
 can be explained as the vev of a scalar field with an odd ``kink'' profile. The vacuum
 configuration of the scalar field resembles in this case the background 3-form field in
 the dimensional reduction of the Horava-Witten theory~\cite{HoravaWitten}.  The $Z_2$-even
 case appears when pseudo-Majorana boundary conditions are imposed, as it will be discussed
 in the following.
\vspace{1truecm}

\noindent {\bf{The odd case}}

\noindent 
A $Z_2$ odd mass parameter is required when ordinary boundary conditions for the bulk 
fermion fields are imposed. These have been extensively considered in the literature~
\cite{GN,GP}. 
The bulk Dirac field transforms under ${Z}_2$ as 
\be
\label{ordinaryBC}
\psi (x, -y)=\pm \gamma_5 \psi (x,y)\, ,
\ee
 where the 
sign can only be determined by the fermion interactions. This transformation property 
implies that the fermion scalar bilinear $\pbp$ is odd under ${Z}_2$, 
$\bar\psi (-y)\psi(-y) = - \bar\psi (y)\psi (y)$. The invariance of the action under $Z_2$ 
 thus requires a ${Z}_2$ odd mass of the type $m(y) = m\,\epsilon (y)$ with 
$\epsilon (y)$ the sign function. 
Given that the left- and right-handed components of the Dirac field satisfy 
$\psi_{L,R} =\mp \gamma_5\psi_{L,R}$, ordinary boundary conditions will imply that the 
L-handed field is ${Z}_2$-odd and the R-handed field even, or viceversa.
The odd component will vanish at the fixed points of the orbifolding 
$y=0$ and $y=\pi R$ - where $\pi R$ is indentified with -$\pi R$\footnote{The orbifolding 
condition that identifies fields at $y=\pi R$ and $y= -\pi R$ is equivalent to periodic 
boundary conditions imposed on universal flat dimensions of size $L$, i.e. 
$\psi (y) = \psi (y+L)$.}.

\noindent The odd bulk mass term vanishes at the boundaries $y=0, \pm \pi R$. 
Thus we conclude that the most general 
mass term allowed by ordinary boundary conditions contains a bulk contribution and no 
boundary terms.

\vspace{1truecm}
\noindent {\bf The even case}

\noindent The ${Z}_2$-even case has instead an even mass paramater $m$ and an even fermion 
bilinear which 
does not vanish at the orbifold fixed points, so that both a bulk mass contribution and a 
boundary contribution are in general allowed.  
We consider an especially suitable type of unordinary boundary conditions, first introduced 
in ref.~\cite{flat} in the context of universal flat extra dimensions. 
We call them {\emph{pseudo-Majorana}} boundary conditions, since they relate the bulk 
fermion field to its charge-conjugate field in the five-dimensional theory and 
they are able to generate the Majorana condition $\psi = \psi^c$ back into four dimensions.

\noindent The pseudo-Majorana boundary conditions are imposed on the Dirac fields living 
in the 5D bulk as follows
\be
\label{eq:tbc}
\psi (x, -y) = e^{-i\delta} \gamma_5\psi^c (x,y)\, ,
\ee
where a phase factor is allowed in the most general case, and $\psi^c =C \left
 ( \bar{\psi}\right )^T$ is the charge conjugate field with 
$C (\equiv C_5)$ the charge conjugation matrix of the five dimensional theory. 
At this point we only need to know the following properties of $C$:
\be
\label{eq:c5}
C^\dagger = C^T = C^{-1} = -C\qquad C\;\Gamma_a \;C^{-1} = \Gamma_a^T\, ,\qquad 
\Gamma_a = (\gamma_\mu, \gamma_5)\, ,
\ee
while in section~(\ref{sec:C5D4D}) we shall state the relation between the 5D charge 
conjugation 
operator and the charge conjugation operator in four dimensions. A valid matrix 
representation of the 5D charge conjugation operator in Dirac or Wyel basis is 
$C=\gamma_1\gamma_3$. It is clear that any boundary condition which induces a mixing of a 
fermion field and its charge conjugate will only be allowed for electrically-neutral fields, 
and this is the case of neutrinos. Any additional $U(1)$ symmetry, under which the fermion
 is charged, will be broken by the boundary conditions in eq.~(\ref{eq:tbc}). 

\noindent In order to make connection to the charge conjugate field, notice that the action 
in eq.~({\ref{eq:action_5d}}) can equivalently be written in terms of the field $\psi$ 
and its charge conjugate in a compact form, also extensively used in ref.~\cite{flat} 
\be
\label{eq:action_bi}
S= \int\, d^4x \int\, dy\, \sqrt{-g}\, \left ( i\bar{\Psi} g^{MN}\gamma_M\,D_N \Psi - 
\bar{\Psi}{\cal{M}}\Psi \right )\, ,   
\ee   
where we have defined the following vector of fields and mass matrix
\be
\Psi\, =\, \pmatrix{\psi^c \cr \psi }\qquad {\cal{M}}\,=\,\pmatrix{-m &0\cr 0 & m } .
\ee
To derive eq.~(\ref{eq:action_bi}) one observes that the action of the 5D charge 
conjugation operator implies $(\psi^c)^c = -\psi$, hence $\pbp = -\psicbar \psi^c$ and 
$\bar{\psi}\gamma^M\,D_M\psi = \bar{\psi^c}\gamma^M\,D_M\psi^c$. These identities
 lead straightforwardly to the form of eq.~(\ref{eq:action_bi}).
The pseudo-Majorana boundary condition on the vector of fields $\Psi$ reads as follows
\be
\label{eq:tbc_compact}
\Psi (x, -y) = \sigma_1\, e^{-i\sigma_3\delta}\,\gamma_5\,\Psi (x,y)\,= 
-\sigma_3\, e^{i\sigma_3\delta}\,\gamma_5\,\Psi^c (x,y) .
\ee
The vector of fields $\Psi (x,y)$ is also subject to periodic boundary conditions, so 
that $\Psi (x, \pi R) = \Psi (x, -\pi R)$. This is a very natural condition to impose. 
However, this type of constraint can also be generalized, see for example~\cite{flat}.

\noindent We perform a 
Kaluza-Klein decomposition of the 5D fields in order to be able to derive 
an action for fermions in four-dimensions. We have the choice of working in terms of
 the fields $\psi$ and $\psi^c$, or in terms of L- and R-handed fields $\psi_{L,R}$,
 or in terms of the
vector of fields $\Psi$.
We choose to decompose the 5D fields in terms of L- and R-handed components to keep our 
derivation easily comparable with the Dirac case of ordinary boundary conditions 
extensively analyzed in the literature~\cite{GN,GP}. We write
\be
\label{eq:KK}
\psi_{L,R} (x, y) = \frac{1}{\sqrt{2\pi R}} \sum_{n=0}^\infty\, \psi_{n,L,R}(x)\, 
e^{2\sigma}\,\hat{f}_{n,L,R}(y)\, ,
\ee
where in the most general case the $y$-dependent functions $\hat{f}_{n,L,R}(y)$ are 
complex valued, and subject to the specific boundary conditions. The same decomposition 
for the charge conjugate field 
$\psi^c(x,y)= C{\bar{\psi}}^T(x,y) = C\gamma_0\psi^\ast (x,y)$ reads
\be
\label{eq:KKC}
\left (\psi_{L,R}\right )^c (x, y) = \frac{1}{\sqrt{2\pi R}} \sum_{n=0}^\infty\, 
\left (\psi_{n,L,R}\right )^c (x)\,  
e^{2\sigma}\,\hat{f}_{n,L,R}^\ast (y)
\ee
and the identity $(\psi^c)_{L,R} = (\psi_{R,L})^c$ holds under the action of the 5D charge
 conjugation operator.

\subsection{ Imposing Pseudo-Majorana boundary conditions}

\noindent 
The boundary conditions in eq.~(\ref{eq:tbc}) constrain the functions $\hat{f}_{n,L,R} (y)$ 
 through the Kaluza-Klein decomposition of eqs.~(\ref{eq:KK}) and (\ref{eq:KKC}), 
so that one 
only independent function remains, associated to one 4D Weyl spinor analogously to what 
happens when imposing the Majorana condition in four dimensions. 
Writing eq.~(\ref{eq:tbc}) in terms of L- and R-handed fields 
we obtain
\ba
e^{i\delta}\psi_L (x,-y) &=& -(\psi^c)_{L}(x,y) =  -(\psi_{R})^c(x,y) \nonumber\\
e^{i\delta}\psi_R (x,-y) &=& +(\psi^c)_{R}(x,y) =  +(\psi_{L})^c(x,y)\, . 
\ea
When deriving the boundary conditions on the KK fields we have the choice to absorb the 
phase factor $e^{i\delta}$ into a) the constraints on the 4D fields $\psi_n(x)$ or into b)
 the constraints on the $y$-dependent functions $\hat{f}_n(y)$. Clearly, both choices are
 equivalent. In the first case one has
\be
\psi_n(x)=e^{-i\delta}\gamma_5 \psi_n^c(x) \nonumber
\ee
and
\ba
\label{eq:4D_BC}
e^{i\delta}\psi_{n,L}(x) &=& - (\psi_n^c)_L(x)= -(\psi_{n,R})^c(x)\nonumber\\
e^{i\delta}\psi_{n,R}(x) &=&  (\psi_n^c)_R(x)= (\psi_{n,L})^c(x)\, ,
\ea
while the functions $\hat{f}_{n,L,R} (y)$ must satisfy the relation
\be
\label{eq:BC_fLR}
\hat{f}_{n,L,R} (-y) = \hat{f}^\ast_{n,R,L} (y)\, ,
\ee
therefore leaving one independent function, as expected.
In the second case, one has $\psi_n(x)=\gamma_5 \psi_n^c(x)$ and $\hat{f}_{n,L,R} (-y) = 
e^{-i\delta}\hat{f}^\ast_{n,R,L} (y)$. 
When instead ordinary boundary conditions are imposed, the Dirac nature of 
the 5D fermion fields is preserved and the two $y$-dependent functions $\hat{f}_{n,L,R}$ are 
subject to the constraints $\hat{f}_{n,L,R} (-y) = \mp \hat{f}_{n,L,R} (y)$, or viceversa.

\noindent In both cases, and after Kaluza-Klein decomposition, the periodic boundary 
conditions $\psi_{L,R}(x,\pi R) = \psi_{L,R}(x,-\pi R)$ lead to $\hat{f}_{n,L,R}(\pi R) = 
\hat{f}_{n,L,R}(-\pi R)$, implying that odd functions of the fifth coordinate
 will vanish at the boundaries $\pm \pi R$\footnote{Notice that in the most general case
 a phase factor is allowed, so that $\psi_{L,R}(x,\pi R) = e^{i\alpha}\psi_{L,R}(x,-\pi R)$.
 Antiperiodic boundary conditions have been considered in ref.~\cite{GP_2} in a 
supersymmetric scenario.}.

\subsection{ Boundary conditions in terms of Weyl spinors.}

\noindent 
We can decompose the 5D Dirac spinor $\psi = \psi_L+\psi_R$ in terms of a L-handed and a
 R-handed Weyl spinor as follows
\be
\psi_L = \pmatrix{\phi \cr 0}\qquad  \psi_R = \pmatrix{0 \cr \epsilon \xi^\ast 
\equiv \chi}\, ,
\ee
where $\epsilon\equiv i\sigma_2$. Charge conjugation in 5D is realized by 
\be
C = \gamma_1\gamma_3 = i\sigma_2 \otimes {\mathbbm{1}} = 
\pmatrix{\epsilon &0 \cr 0 &\epsilon} .
\ee
Working in the Weyl basis with $\gamma^0 = \pmatrix{0 & \mathbbm{1} \cr \mathbbm{1} &0}$ and 
$\gamma_5 = \pmatrix{-\mathbbm{1} & 0 \cr 0 & \mathbbm{1}}$ we obtain 
\be
\psi^c (x,y) = C\gamma^0\psi^\ast = \pmatrix{\epsilon\chi^\ast \cr \epsilon\phi^\ast}
\ee
and the pseudo-Majorana boundary conditions thus imply (in the simplified case $\delta =0$)
\ba
\phi (x,-y) &=& -\epsilon \chi^\ast (x,y) \nonumber\\
\chi (x,-y) &=& \epsilon \phi^\ast (x,y)\, .
\ea  
The boundary condition $\psi (x) =\gamma_5 \psi^c (x)$ on the 4D fields implies
\ba
\phi (x) &=& -\epsilon \chi^\ast (x) \nonumber\\
\chi (x) &=& \epsilon \phi^\ast (x)\, .
\ea  
The 4D spinor can hence be written in terms of one Weyl spinor 
\be
\psi (x) = \pmatrix{ \phi (x) \cr \epsilon \phi^\ast (x)} =  \pmatrix{ -\epsilon 
\chi^\ast (x)\cr \chi (x)}
\ee
under the 5D pseudo-Majorana conditions.

\subsection{The action in four dimensions}

\noindent We determine the spectrum of the KK modes, their masses and wave functions,  
by solving an eigenvalue problem which can be obtained once an ansatz for 
the 4D fermion action is imposed to the 5D action in eq.~(\ref{eq:action_5d}).
The problem is: find the eigenfunctions $\hat{f}_{n,L,R}$ and the associated 
eigenvalues of a $y$-dependent differential operator, subject to the given boundary 
conditions.
 After rewriting the 5D action in terms of L- and R-handed fields
\ba 
\label{eq:action_LR}
S&=& \int\, d^4x \int\, dy\, \sqrt{-g}\, \bar{\psi} (x,y) \left ( i e^\sigma \dslash 
+2\gamma_5 \sigma^\prime -\gamma_5 \partial_5 - m\right ) \psi (x,y)\nonumber\\
&=&\int\, d^4x \int\, dy\, \sqrt{-g}\,\left \{ 
\bar{\psi}_L (x,y)\, i e^\sigma \dslash \psi_L (x,y) + 
\bar{\psi}_R (x,y)\, i e^\sigma \dslash \psi_R (x,y) + \right .\nonumber\\
&+&\left . \bar{\psi}_R (x,y) (-2\sigma^\prime +\partial_5 - m)\psi_L (x,y) +
\bar{\psi}_L (x,y) (  2\sigma^\prime -\partial_5 - m   )\psi_R (x,y) 
\right\} ,      
\ea   
and performing the Kaluza-Klein decomposition according to eq.~(\ref{eq:KK}) 
\ba 
\label{eq:action_KK}
S&=& \int\, d^4x \int\, dy\, \frac{1}{2\pi R}\sum_{n,m}\left \{
 \bar{\psi}_{n,L} (x) i \dslash  \psi_{m,L} (x)\hat{f}_{n,L}^\ast e^\sigma\hat{f}_{m,L}  
+ \bar{\psi}_{n,R} (x) i \dslash  \psi_{m,R} (x)\hat{f}_{n,R}^\ast e^\sigma\hat{f}_{m,R}  
+ \right .\nonumber\\
&+&\left . 
\bar{\psi}_{n,R} (x)\psi_{m,L} (x)\hat{f}_{n,R}^\ast (\partial_5 - m)\hat{f}_{m,L} +
\bar{\psi}_{n,L} (x)\psi_{m,R} (x)\hat{f}_{n,L}^\ast  ( -\partial_5 - m   ) \hat{f}_{m,R}
\right\}\, 
\ea   
we require that the effective 4D action is given by  
\be
\label{eq:action_4D}
S_{4D} = \int\,d^4x\,\sum_n\, \left [\bar{\psi}_n \, i\dslash \psi_n - 
{\cal L}(m_n)\right ]
\, ,
\ee
with a Majorana mass term ${\cal L}(m_n)$. This is guaranteed by the presence of the 
pseudo-Majorana boundary conditions.
The form of the 4D kinetic term in eq.~(\ref{eq:action_4D}) and the use of the
 pseudo-Majorana boundary conditions imply the following 
normalization and orthonormality condition on the $y$-dependent functions
\be
\frac{1}{2\pi R} \int_{-\pi R}^{\pi R}\, dy\, e^\sigma \hat{f}_{n,L,R}^\ast (y)
 \hat{f}_{m,L,R}(y) = \frac{1}{2}\delta_{n,m}\, .
\ee
The most general 4D Majorana mass term involving one Majorana spinor can be 
conveniently expressed as a function of 
L- or R-handed fields and in terms of the 4D charge conjugate fields. This reads 
\be
{\cal L}(m_n) =  m_n ( \psiclbar \psi_L + \bar{\psi}_L \psicl )_n \qquad{\mbox{or}}\qquad 
 m_n ( \psicrbar \psi_R + \bar{\psi}_R \psicr )_n\, ,
\ee
where we defined $\psic_{L,R} \equiv (\psi_{L,R})^c$ and the charge conjugate 
fields are now 
$\psi^c = C_4(\bar{\psi})^T$, where $C_4$ is the charge conjugation matrix in four 
dimensions.

\subsection{Charge conjugation from five to four dimensions}
\label{sec:C5D4D}

\noindent Once the Clifford algebra is built for a given number of spacetime
 dimensions $d=t+s$, with t timelike and s spacelike dimensions, there always exists a
 charge conjugation matrix $C$, such that 
\cite{Tools} 
\be
C^T = -\epsilon\, C\qquad \Gamma_a^T = -\tilde{\eta}\, C\Gamma_a C^{-1}
\ee
for $\epsilon =\pm 1$ and $\tilde{\eta} = \pm 1$, and $\{\Gamma_a, \Gamma_b\} = 
2\eta_{ab}$ defines 
the Clifford algebra in the given spacetime. A formal proof and further details 
can be found in \cite{Tools,Scherk4,Kugo5}.  
The unitarity of the charge conjugation matrix $C^\dagger = C^{-1}$ is true in any 
representation, while the charge conjugation matrix not necessarily squares to $\mathbbm{1}$.

\noindent 
What it is relevant to observe is that for odd dimensions only one of the two choices, 
$\tilde{\eta} = +1$ or $\tilde{\eta} = -1$, can be used \cite{Tools}. While in $d=4$ the 
two solutions 
$\epsilon = +1, \tilde{\eta} = +1$ and  $\epsilon = +1, \tilde{\eta} = -1$ are allowed, 
in $d=5$ only the 
solution $\epsilon = +1, \tilde{\eta} = -1$ can be realized.

\noindent 
An often used realization of the charge conjugation in $d=4$ is $C=i\gamma^2\gamma^0$, which
satisfies
\ba
&&C^\dagger = C^T = C^{-1} = -C\qquad C^2 = -\mathbbm{1} \nonumber\\
&&C\gamma^\mu C^{-1} = -{(\gamma^\mu )}^T\, .
\ea
This choice gives the relation $\gamma_5 C_5 = -C_4$.
In the most general case we can choose a charge conjugation matrix in four dimensions with 
a phase factor, such that $C_4 = \eta\gamma^2\gamma^0$, and $|\eta |^2 = 1$. 
The matrix $C_4$ satisfies the following relations
\ba
&&C^\dagger  = C^{-1} = (\eta^\ast )^2 C\qquad C^2 = \eta^2 \mathbbm{1} \nonumber\\
&& C^T = -C  \nonumber\\
&&C\gamma^\mu C^{-1} = -{(\gamma^\mu )}^T\, ,
\ea
and it is immediate to verify that $\gamma_5 C_5 = -i\eta^\ast C_4$.
In the case in which the phase factor $e^{-i\delta}$ of eq.~(\ref{eq:tbc}) is absorbed
 into the boundary conditions for the 4D fields, they will satisfy the following relation
\be
\label{eq:C5C4}
\psi (x)= e^{-i\delta}\gamma_5 C_5\bar{\psi}^T(x) = -i\eta^\ast e^{-i\delta}C_4 
\bar{\psi}^T(x)=-i\eta^\ast e^{-i\delta}
 \psi^c (x)\, ,
\ee
implying the usual Majorana condition in four dimensions, generalized with a phase
 factor coming from the freedom in the choice of 5D boundary conditions and the 
freedom in the construction of the charge conjugation matrix in four dimensions.
Eq.~(\ref{eq:C5C4}) leads to the identitites 
\ba
\bar{\psi}_R\psi_L &=& -i\eta^\ast e^{-i\delta}\bar{\psi}_R\psi^c_R  = 
i\eta e^{i\delta}\bar{\psi^c_L}\psi_L \nonumber\\
\bar{\psi}_L\psi_R &=& -i\eta^\ast e^{-i\delta} \bar{\psi}_L\psi^c_L  = i\eta e^{i\delta}
\bar{\psi^c_R}\psi_R \, ,
\ea 
thus implying that the 5D action in eq.~(\ref{eq:action_KK}) generates a 4D Majorana mass 
term if subject to pseudo-Majorana boundary conditions.
Equivalently, the phase factor $e^{-i\delta}$ can instead be absorbed into the boundary 
conditions for $\hat{f}_{L,R}(y)$. 

\subsection{ Solutions to the eigenvalue problem}

\noindent In order to obtain the 4D action in eq.~(\ref{eq:action_4D}) the $y$-dependent 
functions $\hat{f}_{n,L,R}$ must be solutions of the equations
\ba
\label{eq:EIGEN}
(-\partial_5 +m) \hat{f}_{n,L}(y) &=& -i\eta^\ast e^{-i\delta}    m_n e^\sigma 
\hat{f}_{n,R}(y)\nonumber\\
(\partial_5 +m) \hat{f}^\ast_{n,R}(y) &=& -i\eta^\ast e^{-i\delta} m_n e^\sigma 
\hat{f}^\ast_{n,L}(y)\, ,
\ea
on the interval $[-\pi R, \pi R]$, with real masses $m, m_n$ and  $\hat{f}_{n,L,R}$ 
complex functions, differently from the ordinary Dirac case. Notice that the $Z_2$
 invariance of the 5D action implies that we can restrict all $y$-dependent 
integrations from $0$ to $\pi R$ and the behaviour of the solutions for negative 
$y$ should be fully determined by their $Z_2$ transformation properties.  
The solutions must also satisfy the pseudo-Majorana boundary conditions
 $\hat{f}_{n,L,R} (-y) = \hat{f}^\ast_{n,R,L} (y)$ and periodic boundary conditions
$\hat{f}_{n,L,R} (-\pi R) = \hat{f}^\ast_{n,L,R} (\pi R)$. 
As anticipated, we can equivalently choose to remove the phase factor 
$e^{-i\delta}$ from equations (\ref{eq:EIGEN}), while imposing the boundary conditions  
$\hat{f}_{n,L,R} (-y) = e^{-i\delta} \hat{f}^\ast_{n,R,L} (y)$.
We choose to illustrate this second case, in which we also set $ -i\eta^\ast = 1$, 
so that $\psi_n (x) = \psi_n^c(x)$ is satisfied. With these substitutions, 
equations (\ref{eq:EIGEN}) should then be 
solved separately for the real and imaginary parts, giving
\ba
\label{eq:EIGEN_Real}
(-\partial_5 +m) Re \hat{f}_{n,L}(y) &=&  m_n e^\sigma Re \hat{f}_{n,R}(y)\nonumber\\
(\partial_5 +m) Re \hat{f}_{n,R}(y) &=& m_n e^\sigma Re \hat{f}_{n,L}(y)\, ,
\ea
and the equations for the imaginary parts are obtained by replacing 
$ Re f_{L,R} \to Im f_{L,R}$.
Notice that eqs.~(\ref{eq:EIGEN_Real}) together with periodic boundary conditions 
tell us that we should seek for solutions with definite and opposite $Z_2$ parity, 
i.e. $Re \hat{f}_{n,L}(y)$ even and $Re \hat{f}_{n,R}(y)$ odd or viceversa. 
This constrains the phase $\delta$ of the boundary conditions and the shape of the bulk 
mass parameter $m$.

\noindent For real $e^{-i\delta}$ the pseudo-Majorana conditions imply that $Re f_L$ and 
$Re f_R$ always carry the same parity (the same being true for the imaginary parts). 
This condition only allows for the zero mode solution of eqs.~(\ref{eq:EIGEN_Real}) to 
survive, thus $m_n =0$, the mass term $m=0$ and $\hat{f}_{0,L}(y)=\hat{f}_{0,R}(y)\sim 
const$. 
For imaginary $e^{-i\delta}$ 
the entire Kaluza-Klain tower is allowed; the pseudo-Majorana boundary condition implies 
in this case  that $Re \hat{f}_L$ and $Im \hat{f}_R$ have the same parity, and opposite to 
$Re \hat{f}_R$ and 
$Im \hat{f}_L$, a condition always compatible with eqs.~(\ref{eq:EIGEN_Real}).

\noindent 
We expect that the complete solutions $\hat{f}_{n,L,R}(y)$ will be plane waves - the 
solutions of the flat case~\cite{flat} - deformed by the warped geometry. Some more 
subtleties arise from the comparison with the ordinary Dirac case as it is shown below.
\vspace{1truecm}

\noindent {\bf The solutions}

\noindent
Solutions to eqs.~(\ref{eq:EIGEN_Real}) can conveniently be found  by deriving a 
second order partial differential equation which decouples the L- and R-handed modes, 
in complete analogy with what has been done in refs.~\cite{GP,GN}. 
The coupled equations (\ref{eq:EIGEN}) are equivalent to the following second order 
partial differential equation on the interval $[-\pi R, \pi R]$
\be
\label{eq:EIGEN_II}
\left ( -e^\sigma\partial_5 e^{-\sigma}\partial_5 + M^2\right ) Re \hat{f}_{n,L,R}(y) = 
m_n^2 e^{2\sigma} Re \hat{f}_{n,L,R}(y)\, ,
\ee
with the mass-like term $M^2$ given by
\be
\label{eq:M2}
M^2= \mp m\sigma^\prime \pm m^\prime +m^2\, ,
\ee
where $\sigma^\prime = k \epsilon (y)$ and the upper (lower) sign for the $L$ ($R$) modes. 
The form of $M^2$ depends on the shape of the bulk mass parameter $m$. 
In the ordinary Dirac case~\cite{GP,GN} the bulk mass must be $Z_2$ odd, thus 
usually defined as $m=c\sigma^\prime$ proportional to the derivative of the warp factor.
In the pseudo-Majorana case the bulk mass is $Z_2$ even, and we can in general allow for 
a bulk and a boundary contribution. However, no further calculation is needed while 
noticing that for $m$ even the mass-like term $M^2$ will carry mixed parity. Hence, for 
any choice of $m$ even, a definite $Z_2$-parity solution to ~(\ref{eq:EIGEN_Real}) 
only exists for $m =0$. This seems to be a genuine consequence of the warped $\mbox{AdS}_5$ 
geometry with the extra dimension compactified on a $S^1/Z_2$ orbifold. We expect that 
some of these constraints might be released for more general choices of the geometry 
of extra dimensions,
for example in the flat case \cite{flat}.  

\noindent It is convenient to rewrite eq.~(\ref{eq:EIGEN_II}) in terms of the new variable 
$x= e^\sigma m_n/k$, and maintain for a few steps more the explicit dependence upon 
the mass-like parameter $M^2$. We obtain
\be
\label{eq:EIGEN_x}
x^2\frac{\partial^2 f_n}{\partial x^2} +\frac{\sigma^{\prime\prime}}{k^2} x 
\frac{\partial f_n}{\partial x} -\left ( \frac{M^2}{k^2} -|x|^2\right ) f_n(x) =0
\, .
\ee
The boundary terms, proportional to $\sigma^{\prime\prime} =  2 k (\delta (y) - \delta 
(y-\pi R))$, should be treated separately and depend on the form of $M^2$. 
Thus we seek for the general solution of
\be
\label{eq:AlmostBessel}
 x^2\frac{\partial^2 f_n}{\partial x^2}  -\left ( \frac{\hat{M}^2}{k^2} -|x|^2\right )
 f_n(x) =0
\, ,
\ee
where the reduced mass-like term $\hat{M}^2$ is obtained by subtracting the boundary terms. 
As noticed in ref.~\cite{GP} a further redefinition $f(x) = e^{\sigma/2} \tilde{f}(x)$
 reduces equation~(\ref{eq:AlmostBessel}) to a Bessel equation for $\tilde{f}_n(x)$
\be
\label{eq:Bessel}
 x^2\frac{\partial^2 \tilde{f}_n}{\partial x^2} + x \frac{\partial \tilde{f}_n}{\partial x}
  -\left ( \frac{\hat{M}^2}{k^2} +\frac{1}{4} -
|x|^2\right ) \tilde{f}_n(x) =0\, 
\ee
the most general solution of which is a linear combination of Bessel functions 
$J_\alpha (x)$ and $Y_\alpha (x)$ of order $\alpha = \sqrt{ {\hat{M}^2}/{k^2} +{1}/{4} }$,
 for $x$ real~\cite{GP,GN}. The complete solutions $f_n(y)$ are thus given by 
\be
\label{eq:solution}
f_n(y) =\frac{e^{\sigma /2}}{N_n}\left [ J_\alpha \left (\frac{m_n}{k}e^\sigma\right ) 
+ b_\alpha (m_n)  Y_\alpha \left (\frac{m_n}{k}e^\sigma\right )          \right ]\, ,
\ee
with normalization factor
\be
N_n^2 = \frac{2}{\pi R}\int_{-\pi R}^{\pi R}\, dy\, e^{2\sigma} \left ( J_\alpha \left
 (\frac{m_n}{k}e^\sigma\right ) + b_\alpha (m_n)  Y_\alpha 
\left (\frac{m_n}{k}e^\sigma\right )      \right )^2\, ,
\ee
where the  coefficients $b_\alpha (m_n)$  and the mass spectrum eigenvalues $m_n$ are
 determined by the specific boundary conditions.

\noindent  When $\hat{M}^2$ has mixed parity, we must solve eq.~(\ref{eq:Bessel}) 
separately on the two branches $y>0$ and $y<0$, thus obtaining Bessel functions of 
different order on the two branches for the same mode $n$. As already concluded before,
 the existence of a solution implies that $M^2 =0$. 
The case $m=0$, $M^2=0$ is immediately solved, the solutions being Bessel functions of
 order $\alpha =1/2$. They are trigonometric functions 
$J_{1/2}(x)=\sqrt{2/\pi x}\, \sin{x}$ and $Y_{1/2}= -J_{-1/2}(x)= -\sqrt{2/\pi x}\,
 \cos{x}$.
The coefficients $b_{1/2}(m_n)$ are determined by the boundary conditions at $y=0, \pi R$ 
(the term proportional to $\sigma^{\prime\prime}$ in eq.~(\ref{eq:EIGEN_x}))
\be
\label{eq:WALL}
{\frac{df_n}{dy}}\vline_{\vspace{0.1cm}y=0, \pi R} = 0
\ee
for $Z_2$ even solutions and 
\be
f_n(y)\vert_{y=0, \pi R} = 0
\ee
for $Z_2$ odd solutions. In this case $f_n(y) = \sigma^\prime /k f_n(|y|)$ and the condition 
$f_n(0)=f_n(\pi R) =0$ directly implies (\ref{eq:WALL}). We obtain the simple result
\ba
\label{eq:bcoeff}
b_{1/2}(m_n) &=& - \frac{ \frac{1}{2} J_{1/2}(\frac{m_n}{k}) + \frac{m_n}{k} 
J_{1/2}^\prime(\frac{m_n}{k})  } { \frac{1}{2} Y_{1/2}(\frac{m_n}{k}) + \frac{m_n}{k}
 Y_{1/2}^\prime(\frac{m_n}{k})  }  = - \cot \left (\frac{m_n}{k}\right )\nonumber\\
b_{1/2}(m_n) &=& b_{1/2}(m_n e^{\pi kR})
\ea
for the even solutions and 
\ba
\label{eq:bcoeff_odd}
b_{1/2}(m_n) &=& - \frac{ J_{1/2}(\frac{m_n}{k}) }{  Y_{1/2}(\frac{m_n}{k}) }  = 
\tan\left (\frac{m_n}{k}\right ) \nonumber\\
b_{1/2}(m_n) &=& b_{1/2}(m_n e^{\pi kR})
\ea
for the odd solutions respectively, with $n=1,2\ldots$. 
Thus the even parity solutions are 
\be
f_n(y) = \sqrt{\frac{k\pi R}{4(e^{k\pi R}-1)}}\, \cos{\left ( \frac{m_n}{k}
\left (e^\sigma -1\right ) \right )}
\ee
and the odd parity solutions are 
\be
f_n(y) = \frac{\sigma^\prime}{k}\sqrt{\frac{k\pi R}{4(e^{k\pi R}-1)}}\, 
\sin{\left ( \frac{m_n}{k}\left (e^\sigma -1\right ) \right )}\, .
\ee
The same conditions, (\ref{eq:bcoeff}) or equivalently (\ref{eq:bcoeff_odd}), give the 
exact spectrum of eigenvalues $m_n$:
\be
m_n =n\pi k \frac{e^{-k\pi R} }{ 1- e^{-k\pi R}}\, .
\ee
This can be compared with the Dirac case with ordinary boundary conditions, where one gets 
an estimate in the limit $kR\gg 1$ and $m_n\ll k$ given by~\cite{GP}  
 $m_n \approx (n+\alpha/2 -3/4) \pi k e^{- k\pi R}$ for even solutions and 
$m_n \approx (n+\alpha/2 -1/4) \pi k e^{- k\pi R}$ for odd solutions, 
and $\alpha =1/2$ in our case.

\noindent Summarizing, the pseudo-Majorana boundary conditions imposed on fermions living on 
a slice of $\mbox{AdS}_5$ generate a Kaluza-Klein tower of Majorana spinors in the four 
dimensional action. No Dirac type bulk mass term seems to be allowed in this case by the 
warped geometry with orbifolding. The phase 
factor $e^{i\delta}$ in the pseudo-Majorana boundary conditions must be chosen along with 
the choice of the 4D charge conjugation matrix. We have derived the 
explicit solutions for pseudo-Majorana boundary conditions given by (imaginary phase) 
$\psi (x, -y) = \pm i\gamma_5 \psi^c (x, y)$, with $C_4= i\gamma^0\gamma^2$. This induces
 the Majorana condition $\psi = \psi^c$ in four dimensions.
A massless zero mode exists and all Kaluza-Klein modes, including the zero mode, are
 localized on the visible ($y=\pi R$) brane. In the same context, we have noticed that 
the opposite 
choice (real phase) $\psi (x, -y) = \pm \gamma_5 \psi^c (x, y)$ removes all excited states
 and only allows for a zero mode. For the moment we leave open the question of what 
mechanism might be in place that favours one choice and not the other.

The $y$-dependent wave functions of the Majorana Kaluza-Klein tower are deformed 
plane waves, as expected, and given by
\ba
\hat{f}_n(y) &=& \sqrt{\frac{k\pi R}{4(e^{k\pi R}-1)}}\, \left ( \cos{
\left (\frac{m_n}{k}\left (e^\sigma -1\right )\right )  } +i \frac{\sigma^\prime}{k}\, 
\sin{\left (\frac{m_n}{k}\left (e^\sigma -1\right )\right )  }  \right )
\qquad n=1,2\ldots \nonumber\\
\hat{f}_0(y) &=& \sqrt{\frac{k\pi R}{4(e^{k\pi R}-1)}}\, .
\ea
These results show that the case of bulk fermion fields with pseudo-Majorana boundary 
conditions is far more constrained than the ordinary Dirac case. 
A bulk Majorana mass term can still be added to our case, in analogy to what has been done 
in the ordinary case in ref.~\cite{HuberShafi}. Again, the mass parameter will acquire 
opposite $Z_2$ parity with respect to the ordinary case if pseudo-Majorana boundary 
conditions are imposed. 
Notice finally that in the ordinary Dirac case the localization of the zero modes is 
monitored by the bulk Dirac mass, which accelerates or reduces the exponential decay of the 
zero mode $y$-dependent wave function. By tuning the bulk mass parameter one can achieve 
localization of the zero modes on the Planck brane, delocalization, or localization on the 
TeV brane. The pseudo-Majorana case produces a zero mode Majorana spinor which is localized 
on the TeV brane.

\section{Neutrino phenomenology, Dirac or Majorana}

The main purpose of this paper is the one of solving the exercise of 
generating Majorana spinors in four dimensions, starting from a five dimensional scenario 
with a warped geometry and bulk Dirac fermions. We did this by imposing non-standard 
pseudo-Majorana boundary conditions in 5D, instead of the ordinary boundary conditions 
which preserve the Dirac nature of bulk fermions. 
This exercise motivates the more phenomenological question of whether the found solution
 can at all give rise to a plausible mechanism to generate small masses for neutrinos.
This is part of the many attempts in the literature at finding a convincing 
explanation of why the observed masses of fermions, quarks and leptons and in particular 
neutrinos, are the way they are, respecting a hierarchy among families and with neutrinos 
much lighter then all their Standard Model partners. 

The pseudo-Majorana boundary conditions do not leave much space for tuning. With this 
mechanism we can generate a zero mode Majorana spinor which is localized on our brane,
 with or without a Kaluza-Klein tower of states.  
We can either identify the bulk fermion field with a ``sterile'' Majorana neutrino,
 i.e. a neutrino which is not charged under the SM gauge group, or with the Standard
 Model neutrino. 
In the first case a well known successful mechanism in four dimensions is the see-saw 
mechanism, where a sterile R-handed heavy neutrino is added to the particle content of 
the Standard Model. 
 We might identify the first Kaluza-Klein mode (or the tower of KK modes) of the Majorana 
spinor in the bulk with the ``heavy'' sterile neutrino and induce a see-saw mechanism, 
while all SM particles are localized on the TeV brane. 
The mass of the Kaluza-Klein
 modes is naturally of the TeV order, and might induce a too small suppression of the SM 
neutrino 
masses. However, a complete description of this mechanism can only be achieved with the full 
understanding of how Kaluza-Klein higher modes are entangled to the four dimensional 
effective theory. This might imply deviations from the assumptions implicit in the 
Kaluza-Klein decomposition. 

\noindent A complete mass term in 4D might be of the type 
\be
S_M = \int~d^4 x \sum_{n\geq 1} m_n \bar{\psi}^{c (n)}_R\psi^{(n)}_R + \sum_{n\geq 0} Y_5
f^{(n)}_R(y=\pi R)\bar{L}(x)<\tilde{H}> \psi^{(n)}_R(x)\, .
\ee 
The usual see-saw mechanism in four dimensions predicts $m_{light} \sim v^2/M$ and
 $m_{heavy} \sim M$, with $v$ the weak-scale vev of the neutral component of the 
Higgs doublet and $M$ a GUT scale. In the 5D case here discussed, and considering 
the tower of Kaluza-Klein excitations we can generate Dirac masses in the four 
dimensional theory through Yukawa couplings 
proportional to $Y_5^{(n)} f^{(n)}_R(y=\pi R) v$ and 
Majorana masses $m_n$ which are naturally of order the TeV. One needs to know the 
behaviour of $Y_5^{(n)} f^{(n)}_R(y=\pi R)$ for the excitations in the KK tower in
 order to predict a 
hierarchy of scales. In the simplest scenario in which all couplings and masses are
 of order the weak scale, no hierarchy will be generated. However, a suppression might
 be induced by the suppression of the couplings of KK excitations to the visible brane.  

\noindent 
As a side note we add that in the case in which Majorana masses and Dirac masses are of the 
same order of magnitude and sufficiently small, the see-saw mass matrix would give rise to 
large mixing angles, inducing neutrino-antineutrino oscillations of the type 
$\nu_L \to (\nu_R)^c$. This would mean 
oscillations in which the SM neutrinos turn into ``sterile'' particles. This alternative 
idea has been explored long ago in \cite{Barger1980,ChengLi1980}.
 
The depicted scenario can be compared with the Dirac case discussed in ref.~\cite{GN}, 
where the R-handed bulk fermion zero mode is identified with a sterile neutrino. 
In this case the bulk mass term can be used as a free parameter 
to tune the amount of localization of the zero mode.
Above a critical value of the bulk mass (c=1/2) the $Z_2$ even (R-handed) zero mode is 
localized on the Planck brane, so that its wave function at the TeV brane is 
strongly suppressed. By introducing a ``new Yukawa'' interaction 
$Y_5 \bar{L}(x)\tilde{H}(x)\psi^{(0)}(x, y=\pi R)$, the four dimensional Yukawa coupling 
will be given by $Y_5 f_{0,R}(y=\pi R)$, with $Y_5=O(1)$ and the SM neutrino mass 
induced via the Higgs mechanism $m_\nu \sim Y_5 vf_{0,R}(y=\pi R)$ receives an 
exponential suppression given by $ f_{0,R}(y=\pi R)$ the value of the $y$-dependent
 function at the TeV brane.
This beautiful mechanism however requires a fine-tuning of the bulk mass term. 
Nevertheless it has been 
shown~\cite{GN} to be very effective in producing plausible ranges of neutrino masses and 
oscillation parameters.

\noindent The second possibility is to identify bulk fermion fields with SM particles.
 This implies that gauge fields should also be 
allowed in the bulk. We do not consider here this 
possibility, rather we observe that a scenario with all SM fields in the bulk offers many 
more appealing features if supersymmetry is imposed. In the simplest case, and disregarding 
gauge interactions, the zero mode of the bulk Majorana field can be identified with the 
SM L-handed neutrino. Given its Majorana nature, no Yukawa coupling to the Higgs doublet and 
therefore Dirac mass term can be generated, while a mass term can be induced by the 
coupling with a Higgs triplet. The smallness of neutrino masses would in this case be 
explained by the smallness of the Higgs triplet vev, a well known mechanism in the four 
dimensional theory. The fitted value of the $\rho$ parameter of the Standard Model
 restricts the ratio of the Higgs triplet to the doublet vevs to be approximately 
$v_T/v_H < 0.17$, thus naturally implying a suppression of neutrino masses with respect to 
charged SM fermions. In this scenario, the question of the smallness of neutrino masses is 
moved into the question of the smallness of a Higgs triplet vev and a hierarchy of 
fermion masses would be due to a hierarchy of scalar boson masses.    

\noindent Alternatively, a mechanism in which higher order radiative corrections give rise 
to a small mass for the Majorana zero mode remains a viable solution.

\section{Other mechanisms to generate small Majorana masses in a warped geometry}

\noindent 
It is now part of textbooks the list of recipes to follow in order to generate a small 
neutrino mass in our four-dimensional world. Three possibilities are foreseen
i) enlarging the SM lepton sector (e.g. with the addition of an $SU(2)$ singlet 
 R-handed neutrino), 
ii) enlarging the SM Higgs sector (e.g. with the addition of a Higgs triplet and 
extra doublets) and iii) enlarging both.
Fermions in the bulk of (compactified) extra dimensions offer an additional mechanism of 
enhancement or suppression of interactions and generation of mass terms. 

\noindent 
However, no clear mechanism has been isolated that can explain why neutrino masses are much 
smaller than charged lepton masses, and most of the proposed mechanisms do rely on a fine 
tuning procedure. Experimentally, no evidence in favour or against the Majorana 
nature of neutrinos has been reported, apart from a long debated claim of the Gran Sasso 
neutrinoless double $\beta$-decay experiment. For an up to date analysis of neutrino 
data see e.g.~\cite{Fogli}. 
Results collected along the years seem to indicate that some relevant 
ingredient is missing, perhaps a fundamental symmetry principle, a group theory 
consideration~\cite{Volkov, Davidson} that might explain the recurrence of the three-family 
structure in the fermion spectrum of the standard model without the need for a fine tuning. 
Might a random process have a role in explaining the observed hierachies? It should 
possibly occur jointly with an enhancement mechanism acting on an initial tiny symmetry
 breaking effect.An intriguing use of randomness has been made in ref.~\cite{Hall} in
 order to describe the apparent hierarchy of athmospheric and solar neutrino oscillation
 parameters. 

\noindent 
In the context of warped extra dimensions, the main attempts to generate a small Majorana 
mass for neutrinos on the visible brane, can be summarized as follows:
i) Higher dimensional operators such as the operator $\psi^TC H^2\psi$~\cite{HuberShafi_2} 
can induce a Majorana mass term on the visible brane. The suppression of the neutrino 
mass w.r.t. 
the charged lepton mass is obtained when the 5D wave function of the zero 
mode is localized on the Planck brane, similarly to what originally proposed in 
ref.~\cite{GN}. ii) The addition of a bulk Majorana mass on the Planck brane~
\cite{HuberShafi} in the case of bulk fermion fields with ordinary boundary conditions
can induce an uplifting of the mass of the bulk fermion zero mode, which can thus 
be identified with the R-handed heavy neutrino of a see-saw mechanism. iii) Finally 
the role of additional scalar fields in the bulk has been explored in ref.~\cite{Chen}.
Assuming a L-handed neutrino on the visible brane, we can generate a Majorana mass term 
via the interaction with an $SU(2)$ Higgs triplet: $\lambda_{ij} L_i^T C^{-1}(i\tau_2 T)
 L_j$, with obvious notation. 
The vev of the neutral component, call it $\xi_0$, of the Higgs triplet $T$ will thus induce 
a Majorana mass term $\sim \lambda_{ii} \xi_0$. Therefore, a suppression of the triplet 
vev w.r.t. the doublet vev would account for a suppression of the Majorana neutrino mass 
w.r.t. the charged fermion masses.
It was concluded in ref.~\cite{Chen} that the introduction of bulk scalar fields would 
suffice to guarantee a suppressed triplet vev on the visible brane, thus offering a viable 
mechanism for the generation of a suppressed Majorana mass for SM neutrinos.
However, after reconsidering the proposed model, we find~\cite{Tesi} that no suppression 
mechanism can actually 
be obtained\footnote{A rescaling of a quartic interaction in going from the 5D to 
the 4D action of the singlet scalar field seems to be missing in ref.~\cite{Chen}}.
We briefly summarize the most relevant steps of the derivation and provide some physics 
considerations.
The model contains one bulk scalar singlet field $S(x,y)$, a Higgs triplet $T(x)$ and a 
Higgs doublet $H(x)$ on the visible brane. Denoting with $\xi_0$ and $v$ the vevs of the
 neutral triplet and doublet respectively, the desired hierachy of vevs should be 
$\xi_0 \ll v$, while the vev of the bulk scalar singlet, call it $S_0$, 
should induce such hierarchy.

\noindent 
The action for a bulk scalar field has also been derived in ref.~\cite{GP}, where the 5D 
equations of motion for bulk scalar, vector and fermion fields are represented in a useful
 compact form. Through power counting, we can write the most general potential for the
 singlet field and its interactions with the triplet and doublet on the brane, up to and 
including quartic couplings. The potential we find does agree with that of ref.~\cite{Chen}.
The bulk singlet field with a generic mass term $ak^2 +b\sigma^{\prime\prime}$ -- a bulk
 and a boundary contributions are allowed by $Z_2$ invariance --   
has a zero mode if $b=2-\alpha$, with $\alpha = \sqrt{4+a}$. This relation between the
 bulk and boundary terms can be justified in the supersymmetric limit~\cite{GP}. 
As a consequence, the localization of the singlet zero mode is driven by the 
parameter $\alpha$: it will be localized on the Planck brane 
for $\alpha > 1$, on the TeV brane for $\alpha <1$ and delocalized for $\alpha =1$.
The author of ref.~\cite{Chen} points out that the Planck localized case can provide 
the required hierarchy.
However, we find a different rescaling of the coupling of the quartic self-interaction of 
the singlet field $\lambda_S$ when reducing from five to four dimensions, with crucial
 consequences for the physics conclusion. 
The coupling $\lambda_S$ rescales as follows
\be
\tilde{\lambda}_S = \frac{\lambda_S}{M_P^2}\frac{1}{4\pi^2 R^2}\frac{1}{N_0^4} 
e^{4(1-\alpha )k\pi R}
\ee
where 
\be
N_0^{-1} = \sqrt {\frac{2(1-\alpha )k\pi R  }{e^{2(1-\alpha )k\pi R} -1  }    }
\ee
and, for $\alpha >1$, $kR\gg 1$
\be
N_0^2 (\alpha >1, kR \gg 1) \simeq \frac{1}{2(\alpha -1)k\pi R}\, .
\ee
For $\alpha >1$ all couplings of the 5D potential rescale as follows
\ba
&& \tilde{\lambda}_H = \lambda_H\qquad \tilde{\lambda}_T = \lambda_T\qquad 
 \tilde{\lambda}_S = \frac{\lambda_S}{M_P^2}k^2 (\alpha -1)^2 e^{-4(\alpha -1)k\pi R}
 \nonumber\\
&&\tilde{\eta} =\eta\qquad \tilde{\xi} = \xi \sqrt{\alpha -1} \sqrt{\frac{k}{M_P}}
 e^{-(\alpha -1)k\pi R}
 \nonumber\\
&&\tilde{\chi}_i =\chi_i ({\alpha -1}) \frac{k}{M_P} e^{-2(\alpha -1)k\pi R}\,\,\, i=1,2
\qquad
\tilde{\mu}_{H,T}^2 =\mu_{H,T}^2k^2 e^{-2k\pi R}\, ,
\ea
where the first line contains all quartic self-interactions, and the remaining two 
lines contain the couplings of the allowed mixed interactions of singlet, triplet and 
doublet, $\eta ,
\xi ,\chi_{1,2}$ and the doublet and triplet masses $\mu_{H,T}$.
For $k\simeq M_P$ the warped geometry is providing a suppression of the couplings. 
Denoting with $A$ the suppression factor 
\ba
A &=& \sqrt{2(\alpha - 1)} \sqrt{\frac{k}{M_P}} e^{(1-\alpha )k\pi R} \qquad \alpha >1 
\nonumber\\
A &=& \sqrt{2(1-\alpha )} \sqrt{\frac{k}{M_P}} \qquad \alpha <1 \nonumber\\
A &=& \frac{1}{\sqrt{kR M_P} } \qquad \alpha =1\, ,
\ea 
we obtain $ \tilde{\lambda}_S =A^4 {\lambda}_S$,
$ \tilde{\chi}_{1,2} =A^2 {\chi}_{1,2}$ and $\tilde{\xi} = A\xi$, thus providing the 
hierarchy of couplings $ \tilde{\lambda}_S \ll  \tilde{\chi}_{1,2} \ll  \tilde{\xi}$.
Once the hierarchy of couplings is obtained, the solutions of the equations which minimize 
the complete potential will provide the hierarchy of the associated vevs.
The set of coupled equations
\ba
&&2\tilde{\lambda}_H v^3 + v (\tilde{\mu}_H^2 +\tilde{\eta}\xi_0^2 +\tilde{\xi} S_0\xi_0
+\tilde{\chi}_2 S_0^2) = 0 \nonumber\\
&&4\tilde{\lambda}_T \xi_0^3 + 2\xi_0 (\tilde{\mu}_T^2 +\tilde{\eta}v^2 +\tilde{\chi}_1 
S_0^2) 
+\tilde{\xi} S_0 v^2 = 0 \nonumber\\
&&4\tilde{\lambda}_S S_0^3 + 2S_0 (\tilde{\chi}_1 \xi_0^2 + \tilde{\chi}_2 v^2) 
+\tilde{\xi} \xi_0 v^2 = 0
\ea
can be easily analyzed without the need for an exact solution.
From the third equation, by assuming $\xi_0\ll v$ we get $\xi_0 \ll AS_0$ and $S_0\sim v/A$,
 thus enhanced! The second equation, using $S_0\sim v/A$, implies that $\xi_0\sim v$ is the 
only allowed non trivial solution.
Therefore, the produced hierarchy is $S_0\sim ve^{(\alpha -1) k\pi R} \gg v$ and 
$\xi_0 \sim v$. 
This conclusion is completely in agreement with the fact that once the singlet field is 
localized on the Planck brane, it has to acquire a vev of order the Planck scale. 
Unfortunately, this also means that no mechanism is left to provide a suppressed triplet 
vev on the visible brane via the addition of a singlet bulk scalar field with 
self-interacting 
potential.
The cases of a delocalized and TeV-localized singlet field are also not relevant to our 
purpose. 
For $\alpha =1$, a power-like enhancement of the singlet field  $S_0\sim \sqrt{kR} v$ is 
generated, while again $\xi_0 \sim v$. For $\alpha <1$ no hierarchy is produced, with 
$S_0\sim  \xi_0 \sim v$.

\section{Conclusions}

In the context of warped extra dimensions, we have explored the possibility of imposing 
boundary conditions on the bulk fermion fields which mix fields and their charge conjugate
 under a $Z_2$ parity transformation involving the warped extra dimension.
We have called these unordinary boundary conditions {\em pseudo-Majorana} conditions, 
since they induce the usual Majorana constraint on spinors in four dimensions. They 
can be written as
\be
\psi (x, -y) = e^{-i\delta} \gamma_5\psi^c (x,y)\, ,
\ee
where a phase factor is allowed in the most general case.
This is a particular case of a general analysis proposed in \cite{flat} in the case of
 universal flat extra dimensions. The warped geometry with orbifolding induces peculiar 
behaviours of the allowed spectrum of fermion fields obtained in  the reduction from five 
to four dimensions via the usual Kaluza Klein decomposition.
       
\noindent In particular, we have shown that the pseudo-Majorana boundary conditions generate
 a Kaluza-Klein tower of Majorana spinors in the four 
dimensional action. No Dirac type bulk mass term seems to be allowed by these boundary 
conditions embedded in the 
warped geometry with orbifolding. The phase 
factor $e^{i\delta}$ in the pseudo-Majorana boundary conditions must be chosen along with 
the choice of the phase in the 4D charge conjugation matrix. 
A massless zero mode exists and all Kaluza-Klein modes, including the zero mode, are
 localized on the visible ($y=\pi R$) brane. We have also noticed that a particular 
choice of the phase factor  $e^{i\delta}$ forbids all excited states and only allows 
for a zero mode solution. 
The $y$-dependent wave functions of the Majorana Kaluza-Klein tower are
plane waves deformed by the warped geometry.

\noindent 
These results show that the case of bulk fermion fields with pseudo-Majorana boundary 
conditions is far more constrained than the ordinary Dirac case. No tuning of a 
localization mechanism via the bulk mass term is allowed in the Majorana case. 
It remains interesting the possibility to generalize these boundary conditions allowing
 for a superposition of $\psi$ and $\psi^c$ after a $Z_2$ transformation.
There is also the option of adding a bulk Majorana mass term, analogously to 
ref.~\cite{HuberShafi}. But again, the mass parameter will acquire 
opposite $Z_2$ parity with respect to the ordinary case if pseudo-Majorana boundary 
conditions are imposed. Bulk Majorana spinors can be naturally identified with sterile 
neutrinos with masses of order the weak scale or with a massless 
Majorana Standard Model neutrino.

\noindent
Finally, after reconsidering a derivation proposed in \cite{Chen}, we have shown that
an additional bulk singlet scalar field cannot induce a suppression of a Higgs triplet 
vev w.r.t. a Higgs doublet vev both confined on the visible brane. 
Unfortunately, our conclusion seems to rule out the possibility that a singlet bulk scalar 
field could induce a small mass for a Standard Model 
Majorana neutrino via the Yukawa interaction with a Higgs triplet. Although we suspect that 
this result extends to more general choices of scalar fields and their interactions, 
an interesting task is still the one of generating small Majorana masses through a 
hierarchy of 
scalar vevs, possibly induced by the geometry of extra dimensions.

\subsection{Acknowledgments}
We aknowledge interesting discussions with Eric Bergshoeff, Mees de Roo, Avihay Kadosh, 
Jos Postma, Duurt Johan van der Hoek, and correpondence with M-C. Chen.

\end{document}